\documentclass{aastex631}
\usepackage{amsmath}
\usepackage{enumitem}
\usepackage{graphicx} 
\usepackage{color}

\begin{document}

\title{Acceleration and Transport of the Unstable Cosmic-ray Isotope $^{60}$Fe in Supernova-Enriched Environments}

%\correspondingauthor{August Muench}
%\email{greg.schwarz@aas.org, gus.muench@aas.org}

\author[0000-0003-1244-172X]{Xin-Yue Shi}
\affil{Department of Astronomy, Nanjing University, 163 Xianlin Avenue, Nanjing 210023, China}
\affiliation{Key Laboratory of Modern Astronomy and Astrophysics, Nanjing University, Ministry of Education, Nanjing, China}
\affiliation{Deutsches Elektronen-Synchrotron DESY, Platanenallee 6, 15738 Zeuthen, Germany}

\author[0000-0001-7861-1707]{Martin Pohl}
\affiliation{Deutsches Elektronen-Synchrotron DESY, Platanenallee 6, 15738 Zeuthen, Germany}
\affiliation{Institute of Physics and Astronomy, University of Potsdam, 14476 Potsdam, Germany}

\author[0000-0001-7761-9766]{Michael M.~Schulreich}
\affiliation{Zentrum für Astronomie und Astrophysik, Technische Universität Berlin, Hardenbergstraße 36, 10623 Berlin, Germany}

\begin{abstract}

The unstable isotope $^{60}$Fe, with a half-life of 2.6 million years, is produced primarily in supernova explosions. The observed presence of $^{60}$Fe in cosmic rays and its detection in deep-sea crusts and sediments suggest two possible scenarios: either the direct acceleration of $^{60}$Fe from supernova ejecta or its enrichment in the circumstellar material surrounding supernova progenitors, which indicates cosmic ray production in clusters of supernovae. Focusing on the latter scenario, we consider an environment shaped by successive supernova explosions, reminiscent of the Local Bubble around the time of the most recent supernova explosion. We independently tracked the evolution of the $^{60}$Fe mass ratio within the Local Bubble using passive scalars. To investigate the spectra of protons and $^{60}$Fe, we explicitly modelled cosmic-ray acceleration and transport at the remnant of the last supernova by simultaneously solving the hydrodynamical equations for the supernova outflow and the transport equations for cosmic rays, scattering turbulence, and large-scale magnetic field, using the time-dependent acceleration code RATPaC. 
The main uncertainty in our prediction of the local 
$^{60}$Fe flux at about $pc=1$~GeV/nuc is the magnetic-field structure in the Local Bubble and the cosmic-ray diffusion beyond the approximately $100$~kyr of evolution covered by our study. 
We found that if the standard galactic propagation applies, the local $^{60}$Fe flux would be around 3\% of that measured. If there is a sustained reduction in the diffusion coefficient at and near the Local Bubble, then the expected $^{60}$Fe flux could be up to 30\% of that measured.

\end{abstract}


\section{Introduction} 
\label{sec:intro}
Supernova remnants (SNRs) are known as promising sites for accelerating particles to relativistic energies, offering persistent targets to study cosmic-ray (CR) acceleration across various wavebands and messengers. 
In recent years, observations in the GeV-TeV energy range and CR detectors have accumulated extensive data from SNRs, providing essential constraints for theoretical models.
The mechanism capable of accelerating CRs to such high energies is widely believed to be diffusive shock acceleration (DSA) \citep{1949PhRv...75.1169F,1978MNRAS.182..147B,1983RPPh...46..973D}.

In the scenario of core-collapse supernovae, the progenitors are massive stars, which would generate complexly modified wind-blown bubbles in the environment\citep{2003ApJ...594..888F, 2014ASTRP...1...61M,2020MNRAS.493.3548M}.  Particle acceleration and subsequent emission processes \citep{2013HEDP....9..226D} are influenced by the interactions between the SNR and its environment.
Particle acceleration during the expansion of an SNR through a stellar wind bubble was studied by \citealt{1988ApJ...333L..65V} and \citealt{2000A&A...357..283B} in the Bohm diffusion regime.
The impact of mass-loss rates during stellar evolution on particle acceleration has been investigated using simplified flow profiles by \citealt{2022ApJ...926..140S} and \citealt{2023JOce...79..379K}.
The spectral evolution and emission morphology of SNR interacting with realistic circumstellar material (CSM) were studied for both Bohm-like diffusion and diffusion in self-generated magnetic turbulence \citep{2022A&A...661A.128D,2024A&A...689A...9D}.

The conditions at the acceleration sites and the environments through which these CRs propagate in SNRs still remain under debates \citep{2012APh....39...52D}.
Unstable isotopes serve as crucial tracers in unraveling the mystery of CR transport processes. One such isotope, iron-60 ($^{60}$Fe), is unstable and only produced in supernovae.
With a relatively long half-life of approximately 2.62 million years \citep{2009PhRvL.103g2502R, 2015PhRvL.114d1101W}, $^{60}$Fe not only survives long enough to travel significant distances but also can be detected via its $\beta$-decay to the daughter isotope $^{60}$Co, which results in gamma-ray emission at 1.17 and 1.33 MeV from the activated $^{60}$Co nucleus.
The idea of using $^{60}$Fe and other long-lived radioisotopes (with half-lives of the order of millions of years) as tracers for near-Earth supernova activity was first proposed by \cite{1996ApJ...470.1227E} and \cite{korschinek199660fe}, and has since been extensively investigated in both theory and observation \citep{2007A&A...469.1005W, 2021ApJ...923..219W, 2023EPJA...59...52K, 2024NatAs...8..983O}.

Measurements of CRs with energies from approximately 195 to 500 MeV/nucleon show a source ratio of $^{60}\text{Fe}/^{56}\text{Fe} = (7.5 \pm 2.9) \times 10^{-5}$ \citep{2016Sci...352..677B}, indicating that the $^{60}$Fe sources must have been within a distance that high-energy particles can travel over the duration of its half-life, typically less than about 1 kpc.
This observational result is consistent with theoretical predictions from numerical models \citep{2000SSRv...92..133M, 2018ApJ...863...86B}, which also obtain a similar $^{60}$Fe/$^{56}$Fe source ratio.

Moreover, $^{60}$Fe have been discovered in deep-sea ferromanganese (FeMn) crusts, nodules, sediments (including fossilized bacteria), lunar soil and various locations \citep{2016PNAS..113.9232L, 2016Natur.532...69W, 2021Sci...372..742W, 2019PhRvL.123g2701K, 2016PhRvL.116o1104F}. Geological studies of variations in $^{60}$Fe concentrations in deep-sea sediments further support the interpretation that nearby core-collapse supernovae are the primary sources of CR $^{60}$Fe \citep{1999PhRvL..83...18K, 2025ApJ...979L..18N}.
In particular, the Scorpius–Centaurus OB association has been identified as the most probable origin of the observed $^{60}$Fe nuclei \citep{2002PhRvL..88h1101B, 2006MNRAS.373..993F, 2024MNRAS.530..684D}.

This observed presence of $^{60}$Fe among CRs suggests its acceleration either from supernova ejecta or an enrichment of $^{60}$Fe in the CSM surrounding the supernova progenitor, implying CR production in clusters of supernovae. 
In this study, we focus on the latter scenario: the acceleration of $^{60}$Fe at the forward shock of a core-collapse supernova in an environment enriched with $^{60}$Fe produced by previous supernovae.

The Solar System has been traveling through the Local Bubble (LB)—a low-density cavity in the interstellar medium shaped by stellar winds and supernovae—for the last five to ten million years (see the review by \citealt{2011ARA&A..49..237F}). Recent observations reveal that many nearby star-forming regions lie on the surface of the LB, with young stars expanding perpendicularly from it, suggesting that its expansion may have triggered local star formation \citep{2022Natur.601..334Z}.
Measurements of Galactic cosmic-ray (GCR) source composition show that GCRs preferentially originate from regions of the clustering of massive stars and supernova explosions in the superbubbles \citep{1997ApJ...487..197E, 2004A&A...424..747P, 2021MNRAS.508.1321T}.

In parallel, previous studies have simulated the inflow of unstable isotopes from past supernova activity, showing that the same feedback responsible for forming the LB can also account for the observed enrichment of the local interstellar medium \citep{2017A&A...604A..81S,2023A&A...680A..39S,2023ApJ...944..121W}.
These simulations successfully explain the overabundance of $^{60}$Fe in the time range 1.7--3.2\,Myr ago measured at Earth in deep-sea FeMn crusts and nodules, in ocean sediments from all major oceans, and in the fossilized remains of magnetotactic bacteria, which incorporate iron into their cellular structures \citep{1999PhRvL..83...18K, 2004PhRvL..93q1103K, 2008PhRvL.101l1101F, 2016PNAS..113.9232L, 2016Natur.532...69W, 2021Sci...372..742W}. For an earlier $^{60}$Fe signal from 6.5--8.7\,Myr ago, which was also measured, they suggest the passage of the Solar System through a neighboring interstellar cavity.

In the LB, a number of unstable isotopes, including $^{60}$Fe, are released by supernova explosions after their progenitor stars reach the end of their lifetimes, in accordance with stellar evolution models \citep{2012A&A...537A.146E, 2018ApJS..237...13L}. %\citep{2006ApJ...647..483L, 2007PhR...442..269W}.
These radioisotopes are modelled as decaying passive tracers, implying that their abundances are so low relative to the total density that they do not influence the fluid dynamics but are transported by the fluid flow according to an advection-diffusion equation.
Consequently, the density profile of $^{60}$Fe differs from the overall density.
The bubble region, characterized by a relatively low density, a high temperature, and a higher mass ratio of $^{60}$Fe comparing to the dense shell, represents a unique environment for studying CR acceleration. 
The differences in mass, charge, and particularly the mass fraction of $^{60}$Fe lead to differing injection rates at various stages, resulting in distinct spectral features.

This study aims to investigate the particle spectrum of $^{60}$Fe and compare it with the proton spectrum, providing insights into the CR acceleration mechanisms within SNRs. Using our own particle acceleration code, RATPaC (Radiation Acceleration Transport Parallel Code; \citealt{2012APh....35..300T, 2013A&A...552A.102T, 2018A&A...618A.155S, 2020A&A...634A..59B}), we solved the CR transport equation and hydrodynamics together in 1-D spherical symmetry. As supernova outflows are highly supersonic, they can be approximated as conical segments that evolve quasi-independently. Treating a statistically meaningful set of such segments with RATPaC, we can model the entire SNR. We combine
RATPaC with the PLUTO code, that would calculate the hydrodynamical evolution of the outflow on the fly, and can in this way trace the evolution of supernova forward shock inside the complex LB simulated earlier \citep{2017A&A...604A..81S,2023A&A...680A..39S}.
The mass ratio of $^{60}$Fe is tracked using passive tracers throughout the simulation of the SNR. By analyzing these spectra under different magnetic-field assumptions, we aim to further understand the role of supernovae in CR production and the complex processes that govern their acceleration.

The paper is arranged as follows. Section \ref{sec:method} describe the numerical methods in this study. Section \ref{sec:results} presents the time evolution of the forward shock in the SNR and the resulting $^{60}$Fe and proton spectra. We discuss the ratio of $^{60}$Fe to normal iron, scaled to the local cosmic-ray flux, and compared it with the measured flux in Section \ref{sec:dis}. We present our conclusions in Section \ref{sec:con}.

\section{Numerical methods}
\label{sec:method}
This section introduces the numerical setups employed in the study.
The diffusive shock acceleration at the SNR forward shock was modelled in test-particle approximation by combining the hydrodynamic evolution of the SNR, the large-scale magnetic field profile and the solution for the CR transport equation. The particle acceleration and hydrodynamics were numerically solved with RATPaC and the PLUTO code (\citealt{2007ApJS..170..228M, 2018ApJ...865..144V}) on the fly.
The details of the hydrodynamics, particle transport equation and the modeling of physical parameters of the SNR and its ambient environment are presented below.

\subsection{Hydrodynamics}
The evolution of a SNR can be described with the Euler hydrodynamic equations,

\begin{align}
\label{eq:hd}
\frac{\partial}{\partial t} \left(
\begin{array}{c}
\rho \\ {\bf m} \\ E
\end{array}
\right)
+ \nabla \cdot \left(
\begin{array}{c}
\rho {\bf u} \\ {\bf m u} + P {\bf I} \\ (E+P) {\bf u}
\end{array}
\right)^T
= \left(
\begin{array}{c}
0 \\ 0 \\ 0
\end{array}
\right)
\end{align}
where $\rho$ is the mass density, ${\bf u}$ is the velocity, ${\bf m} = \rho {\bf u}$ is the momentum density, $P$ is the thermal pressure, 
%$L$ the energy losses due to cooling 
and $E$ is the total energy density:
\begin{equation}
E = \frac{\rho {\bf u}^2}{2} + \frac{P}{\gamma -1},
\end{equation}
$\gamma = 5/3$ for ideal gas.
{\bf I} is the unit tensor.
Note that, the simulations are calculated in PLUTO with pure hydrodynamics based on the assumption that the magnetic field is not dynamically significant.

Furthermore, a passive scalar was introduced to track the mass ratio  of $^{60}$Fe to the total density. This reflects that at such low concentrations, the flow is not affected, so the $^{60}$Fe are carried by the fluid according to an simple advection equation of the form
\begin{equation}
\frac{\partial \rho_\mathrm{60Fe}}{\partial t} + \nabla \cdot (\rho_\mathrm{60Fe}{\bf u}) = 0.
\end{equation}

\subsubsection{Ambient environment of the supernova}
\label{sec:ambient}

To accurately simulate the supernova explosion within the ambient medium, it is essential to first understand the nature of the ambient medium, the LB, which is one of many cavities that exist in the interstellar medium (ISM) of our Milky Way, as well as in other star-forming galaxies. These cavities are filled with hot plasma and surrounded by shells of cold, dusty gas.

\citet{2023A&A...680A..39S} conducted high-resolution three-dimensional hydrodynamic simulations to explore the evolution of the LB within an inhomogeneous local interstellar medium and to study the transport of radioisotopes to Earth. Their simulations utilized initial conditions that account for the effects of 14 identified SN explosions occurring within subgroups of the Scorpius-Centaurus OB association, based on data from \emph{Gaia} EDR3. The study also employed a Monte Carlo-type method to determine the trajectories of the supernova progenitor stars, set up their winds depending on their individual ages and initial masses, and considered the ejection of the radioisotopes $^{60}$Fe, $^{26}$Al, and $^{53}$Mn. Furthermore, the dispersion of $^{244}$Pu was investigated, which was assumed to be pre-seeded, possibly by a kilonova event before the formation of the LB.
These simulations were carried out in a cubic computational domain with an edge length of 800\,pc in which a maximum resolution of 0.781\,pc was achieved by means of adaptive mesh refinement.

\begin{figure}
    \centering
    \includegraphics[width=0.9\linewidth]{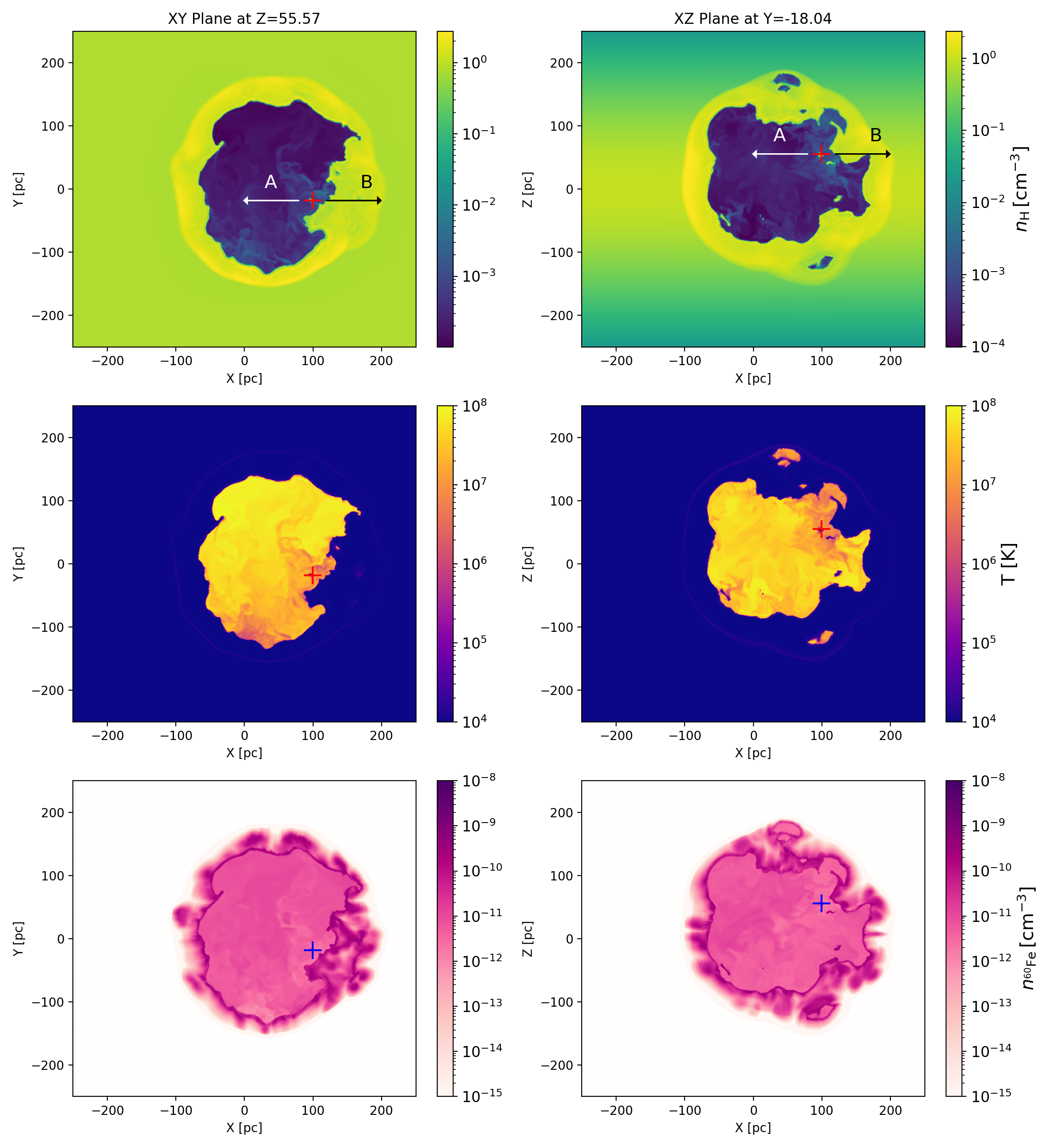}
    \caption{2D axis-aligned slices in the XY and XZ planes through the 3D computational domain, zoomed in on the central 500 pc region of the full 800 pc simulation box, showing (from top to bottom) the hydrogen number density, temperature, and $^{60}$Fe number density in the Local Bubble. The time is approximately 1.68 Myr before present, just prior to the supernova for which we model particle acceleration and whose location is marked by a plus sign. Two representative directions, labeled A (white) and B (black), are indicated in the top panel.}
    \label{fig:slice}
\end{figure}

We use a snapshot taken approximately 1.68 Myr before present, just prior to the penultimate supernova explosion from a 13.28 M$_\odot$ progenitor. This explosion is the one whose remnant we model to investigate particle acceleration. The snapshot provides the initial conditions for our simulation of $^{60}$Fe acceleration, including the large-scale structure of the ambient medium and the spatial distribution of $^{60}$Fe.
At the beginning of our simulation, the cumulative amount of $^{60}$Fe within the simulation domain is about $5 \times 10^{-4}$ M$\odot$, after accounting for radioactive decay. Notably, almost 80\% of the present-day $^{60}$Fe originates from earlier events prior to the modelled supernova.

Figure \ref{fig:slice} shows 2D axis-aligned slices of the snapshot at the position of the SN.
Given that RATPaC assumes spherical symmetry, and the SN environment is not spherically symmetric, we accounted for anisotropy by selecting ten nearly uniformly distributed directions from the location of the supernova. 

We introduce two typical directions out of the ten studied directions and illustrate them as direction A and B in Figure \ref{fig:slice}.
Direction A corresponds to the vector (-1,0,0), which is aligned along the negative x-axis. As shown in Fig.\ref{fig:slice}, this direction points toward the bubble central region, where the forward shock is expected to propagate through an extended region of $\approx 150$ pc filled with hot gas of low density.
Direction B, corresponding to the vector (1,0,0), is aligned along the positive x-axis. This direction was selected because along this path the SNR shock is expected to quickly encounter the edge of the bubble, reaching the dense shell at a distance of $\approx$20 pc from the location of the supernova.

Then we performed ten separate 1D spherically symmetric simulations of CR acceleration along these directions, using RATPaC. The final results were obtained by integrating the data from these simulations across the directions, allowing for a comprehensive analysis of directional effects. The typical flow velocities in the LB are 20–150 km/s, according to hydrodynamical model. The radial components enter our simulations and are shown in Figures 2 and 3. The tangential velocity components must be ignored in the spherically symmetric setup of the simulation.

\subsubsection{Supernova ejecta profile}
A common assumption for the density distribution of supernova ejecta is a constant density $\rho_\mathrm{c}$ up to $r_\mathrm{c}$, then followed by a power law with index $n$ to the ejecta radius $R_\mathrm{ej}$ \citep{1982ApJ...258..790C},

\begin{align}
\rho_\mathrm{ej} (r) = \left\{
\begin{array}{lc}
\rho_c  & r \leq r_c \\
\rho_c (\frac{r}{r_c})^{-n} \ \ \ \ & r_c < r \leq R_\mathrm{ej}
\end{array}
\right.
\end{align}
with
\begin{equation}
r_\mathrm{c} = \left(\frac{10E_\mathrm{ej}}{3M_\mathrm{ej}} \frac{n-5}{n-3} \frac{n-3x^{3-n}}{n-5x^{5-n}}\right)^{1/2} t_\mathrm{SN},
\end{equation}
\begin{equation}
\rho_\mathrm{c} = \frac{M_\mathrm{ej}}{4\pi r_c^3} \frac{3(n-3)}{n^2} (n-3x^{3-n})^{-1},
\end{equation}
where $\rho_\mathrm{c}$ is the density of the inner plateau, $r_\mathrm{c}$ is the radius of the transition from the constant density region to the power-law part, and $n$ the power-law index of the outer ejecta ($n=9$ when considering core-collapse supernovae). $M_\mathrm{ej}$ is the ejecta mass, resulting from an SN explosion with an initial progenitor star mass of 13.28 $M_\odot$, which is the penultimate SN explosion in the scenario described by \cite{2023A&A...680A..39S}, leading to better agreement with the $^{60}$Fe measurements. To maintain consistency, we adopt their wind-loss model for the progenitor star, resulting in an ejecta mass of $M_\mathrm{ej} = 8.32 M_\odot$.
$E_\mathrm{ej} = 10^{51}$ erg is the explosion energy and $R_\mathrm{ej} = x r_c$ is the outer radius of the ejecta, $x=2.5$.
$t_\mathrm{SN} = 3$ yr is the age of the remnant which also represents the starting time of the hydrodynamic simulation.

The velocity profile for ejecta follows homologous expansion, described as
\begin{equation}
v_\mathrm{ej} = \frac{r}{t_\mathrm{SN}}.
\end{equation}
The initial ejecta temperature is set to $10^4$ K, and the pressure is set according to this temperature.

In order to initiate the supernova explosions, the pre-calculated fluid profile within the $R_\mathrm{ej}$ region is replaced with the supernova ejecta profiles. Subsequently, Eq.\ref{eq:hd} is solved with radiative cooling using a Harten-Lax-Van Leer approximate Riemann solver that employs the middle contact discontinuity (hllc), finite-volume methodology, and a third-order Runge-Kutta (RK3) method. The numerical simulations of hydrodynamics were conducted using the PLUTO code in one-dimensional (1-D) spherical symmetry, with a further interpolation to a fine spatial resolution of 0.0003 pc.

\subsection{Cosmic-ray transport equation}
The transport equation for the differential number density of particles of momentum $p$ at location $r$ and time $t$, $N(p,r,t)$, can be expressed in the test-particle limit as
\begin{equation}
\frac{\partial N(p,r,t)}{\partial t} = \nabla \cdot (D \nabla N -{\bf u}N) - \frac{\partial}{\partial p}\left[(\dot{p} N) - \frac{\nabla \cdot {\bf u}}{3} Np\right]  + Q(p,r,t),
\end{equation}
where $D$ is the spatial diffusion coefficient, which will be described in Section \ref{sec:coeD}, $\dot{p}$ is the energy loss rate (e.g. synchrotron and inverse Compton losses for electrons) and $Q$ is the source term.

This equation has been solved simultaneously with hydrodynamics in one-dimensional (1D) spherical symmetry in RATPaC, applying implicit finite-difference algorithms implemented in the FiPy package \citep{2009CSE....11c...6G}.
The equation is transformed to co-moving, shock-centred coordinates to account for the expansion of the supernova, with $x = r/R_\mathrm{fs}$, where $R_\mathrm{fs}$ is the forward-shock radius. Furthermore, in order to achieve better spatial resolution close to the shock, we employed a transformation of the radial coordinate, expressed as $(x-1) = (x_* -1)^3$, for the specific configuration employed in this study, with the resolution in our simulation is set to $\Delta x_* = 0.0125$, yielding $\Delta x \approx 2\times 10^{-6}$ at the shock.
In real space, the inhomogeneous grid extends out to several tens of forward-shock radii in the upstream region, ensuring that all injected particles remain within the simulation domain.

In the simulation, the CR pressure is always below 10\% of the shock ram pressure. This indicates that the impact of the non-linear shock modification by CR pressure is minimal and can be disregarded \citep{2010ApJ...721..886K}. We also note that for the evolution of the system beyond the time covered by our simulation, the CR pressure in the LB is negligible, on account of fast diffusive transport and hence dilution of the CR contribution of each supernova in the LB.

\subsubsection{Particle injection}
We use the thermal leakage model \citep{PhysRevE.58.4911,2005MNRAS.361..907B} for the injection of particles, which is given by
\begin{equation}
Q = \eta N_\mathrm{u} (v_\mathrm{sh} - v_\mathrm{u}) \delta(r-R_\mathrm{fs})\delta(p-p_\mathrm{inj})
\end{equation}
where $N_\mathrm{u}$ is the plasma number density in the upstream region, $v_\mathrm{sh}$ is the shock speed in the simulation, and $v_\mathrm{u}$ represents plasma speed in the upstream region.

Protons are injected with momentum $p_{\mathrm{inj},p} = \xi p_\mathrm{th} = \xi \sqrt{2m_p k_B T_\mathrm{d}}$ at the position of the forward shock, where $T_\mathrm{d}$ is the downstream temperature.
The total injected momentum of $^{60}$Fe is determined by the gyroradius, under the assumption that $^{60}$Fe with the same gyroradius as the proton has equal probability to cross the shock front. Then, $p_\mathrm{inj,60fe} = Z p_{\mathrm{inj},p}$, where Z = 26 is the charge number of $^{60}$Fe.
The injection efficiency is defined as 
\begin{equation}
\eta = \frac{4}{3\sqrt{\pi}} (\chi -1) \xi^3 e^{-\xi^2},
\label{eq:inj}
\end{equation}
where $\chi$ represents the compression ratio of each shock.
We have used $\xi =4.2$ in our simulations, that is consistent with the observed radiation flux from SN1006 \citep{2021A&A...654A.139B} and conforms with the test-particle limit.
Furthermore, based on the observation that heavy elements are preferentially injected into the acceleration process \citep{2007ARNPS..57..285S, 2019ApJS..245...30L}, we chose to apply the same injection efficiency to both $^{60}$Fe and protons. The iron-to-proton ratio will be discussed in more details in section \ref{sec:dis}.

\subsection{Magnetic field}
The total magnetic field strength is given by
\begin{equation}
    B_\mathrm{tot} = \sqrt{B^2 +\langle \delta B^2 \rangle} ,
\end{equation}

where $B$ is the large-scale magnetic field and $\delta B$ is the magnetic turbulence.

\subsubsection{Large-scale magnetic field profile}
\label{sec:B}
In the simulations, the large-scale magnetic field is assumed to be fully randomized, consisting of one radial and two tangential components. This is a far simpler structure than what actual polarization data suggest \citep{2020A&A...636A..17P,2025A&A...694A..97P}, but one that eliminates a model dependence. Faraday-rotation may be dominated by features in the cavity walls \citep{2024A&A...688A.200E}, and the interpretation for the interior of the LB depends on the assumptions on the abundance of ionized gas at the cavity walls.

For the normalization of the initial magnetic field in the upstream region, we adopted two different assumptions. 
The first assumption is a density-coupled large-scale magnetic field, i.e., $B(r) \propto n_\mathrm{H}^{2/3}$, where the magnetic field is normalized to $B = 10\ \mu \mathrm{G}$ at the density of $n_\mathrm{H} = 1\ \mathrm{cm^{-3}}$.
The second assumption is a pressure-coupled magnetic field, i.e., $B(r) \propto P(r)$, with the magnetic field normalized to $B = 10\ \mu \mathrm{G}$ at the thermal pressure of $P = 10^{-12}\ \mathrm{dyne\,cm^{-2}}$. Compared to the polarization-based measurements, with density scaling the magnetic-field strength in the interior of the LB is on the low side of what the data suggest, whereas with pressure scaling it is on the high side.
As the fluid speed in the upstream region remains relatively low, with $u \lesssim 100$ km/s throughout the simulation time of $\sim 20-100$ kyr, the resulting advection is on the order of a few parsecs and can therefore be neglected. Thus, we assume that the magnetic field remains unchanged in the upstream region.

The tangential components of the magnetic field experience compression at the shock, while the radial component remains uncompressed.
Then, in the downstream region, we solved the passive transport equation of the magnetic field
\begin{equation}
    \frac{\partial {\bf B}}{\partial t} = \nabla \times ({\bf u} \times {\bf B}),
\end{equation}
following \citealt{2013A&A...552A.102T}. This method mimics MHD for negligible magnetic pressure.

In the ejecta region, to ensure that Gauss’ law ($\nabla \cdot {\bf B} =0$) is initially satisfied, we assume that $B(r) \propto 1/r^2$ with an inner plateau for both radial and the tangential components.

\subsubsection{Magnetic turbulence}

The transport of magnetic turbulence can be described by a continuity equation for the spectral energy density, $E_w = E_w(r, k, t)$:
\begin{equation}
    \frac{\partial E_w}{\partial t} 
    = - \nabla \cdot ({\bf u} E_w)
    - k\frac{\partial}{\partial k} \left(k^2 D_k \frac{\partial}{\partial k} \frac{E_w}{k^3}\right)
    +2(\Gamma_g -\Gamma_d) E_w,
\end{equation}
where $k$ denotes the wave-number, $D_k$ is the diffusion coefficient in wave-number space describing cascading, $\Gamma_g$ and $\Gamma_d$ are growth and damping rates, respectively.
This transport equation has been solved in 1D spherical symmetry alongside the equation for the CRs and the large-scale magnetic field.

We only consider resonant interactions between waves and CRs. Alfvén waves generated by particles should have wavelengths similar to the gyro-radii of the particles, then the resonance condition is
\begin{equation}
    k_\mathrm{res} = \frac{qB}{pc},
\end{equation}
where $q$ is the particle charge. This condition provides the link between $p$ and $k$,
so the growth rate due to the resonant amplification can be expressed as
\begin{equation}
    \Gamma_g = A \frac{v_A p^2 v}{3 E_w} \left| \frac{\partial N_p}{\partial r} \right| ,
\end{equation}
where $A$ is a linear scaling factor, $v_A$ is the Alfvén speed, $N_p$ is the differential number density of CR protons, which means that the turbulence is generated only by CR protons in the simulation domain.

In this study, we have adopted $A = 10$ to enhance the growth rate of the resonant streaming instability, supporting efficient amplification of the turbulent magnetic field. This is consistent with observations of historical SNRs \citep{2021A&A...654A.139B}.
The damping rate due to neutral-charged collisions and ion-cyclotron are derived from \citealt{2016A&A...593A..20B}.

The spectral energy transfer process from small
wave-number scale to large wave-number scale through turbulence cascading can be described as a diffusion process in wave-number space with coefficient
\begin{equation}
    D_k = k^3 v_A \sqrt{\frac{E_w}{2B^2}}.
\end{equation}
In the case of pure cascading from large scales, this phenomenological treatment will result in a Kolmogorov-like turbulence spectrum, $E_w \propto k^{-2/3}$.

\subsubsection{Spatial diffusion coefficients}\label{sec:coeD}

The diffusion coefficient directly influences the acceleration timescale and the maximum energy of the particles.
In the scenario of the self-consistent treatment for the diffusion coefficient with magnetic field turbulence, we assume an initial seed turbulence with
\begin{equation}
D = \zeta D_0 \left(\frac{pc}{10 \mathrm{GeV}}\right)^{\alpha}\left(\frac{B}{3 \mu\mathrm{G}}\right)^{-\alpha},
\end{equation}
where $\alpha = 1/3$, and $\zeta D_0 = 10^{27}\ \mathrm{cm^2 s^{-1}}$ which is a factor of 100 lower than that for Galactic propagation of CRs \citep{2011ApJ...729..106T}.
The diﬀusion coeﬃcient of a particle with momentum $p$ moving in the background magnetic field with energy density $U_B$ can be calculated using
\begin{equation}
    D_r = \frac{4v}{3 \pi} r_g \frac{U_B}{E_w}.
\end{equation}
where $r_g$ represents the gyro-radius of the particle.

\section{Results}
\label{sec:results}

As mentioned before in section \ref{sec:ambient}, to account for the inhomogeneous and anisotropic environment, we selected ten relatively uniformly distributed directions from the supernova location and conducted separate 1D simulations for each. The final results presented here are derived by integrating the data from these directions.

In this section, we present the respective results from direction A and B in Figure \ref{fig:slice}, showing the time evolution of the forward shock and the particle momentum spectrum for both $^{60}$Fe and protons. 
Subsequently, we will present the integrated final results.

\subsection{Initial HD profile and time evolution of the forward shock}

\begin{figure}
    \centering
    \includegraphics[width=1.0\linewidth]{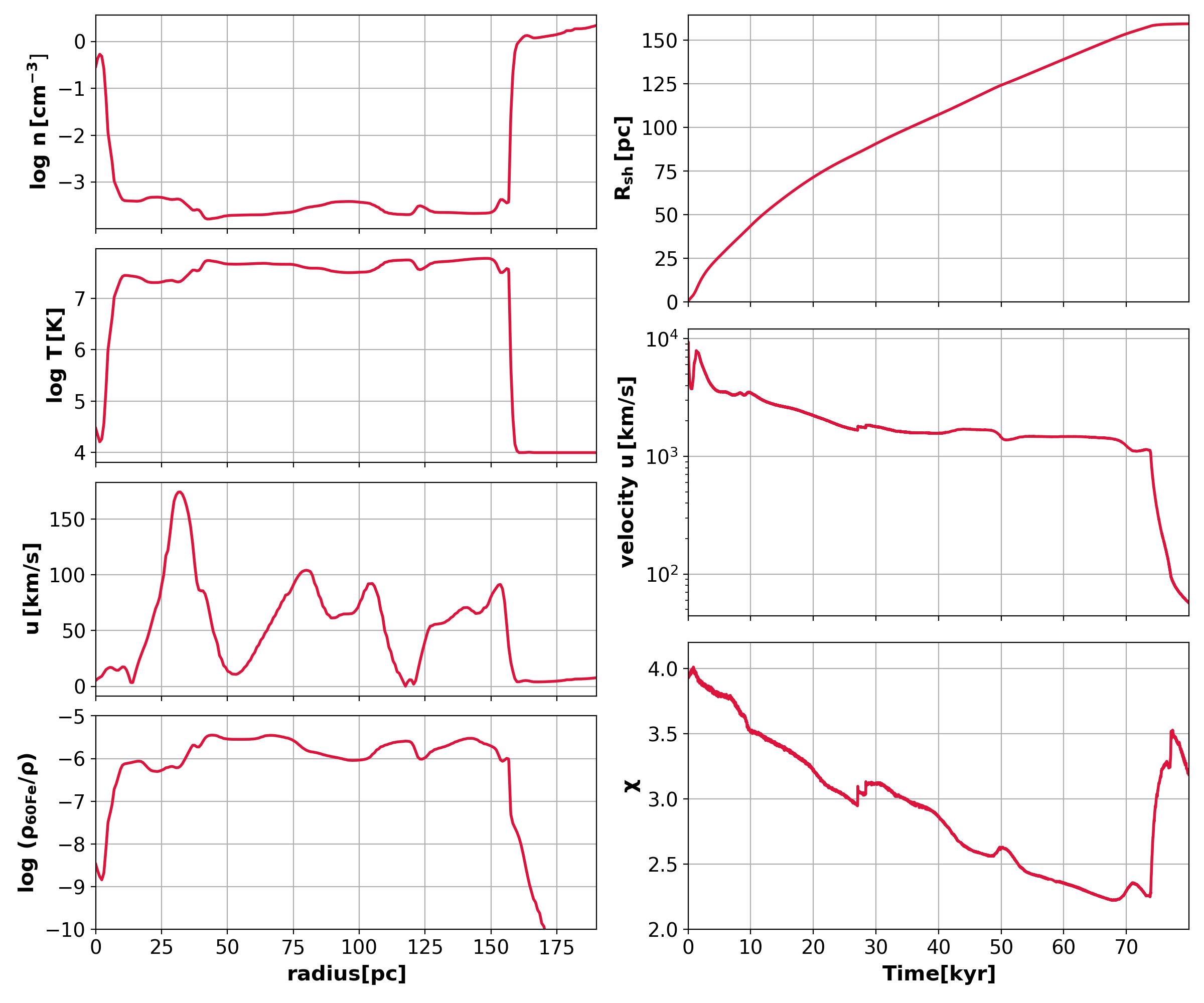}
    \caption{LEFT: Initial hydrodynamic profile of the LB at the explosion time of the modelled supernova, including the gas number density, the flow speed, the temperature, and the $^{60}$Fe mass fraction in direction A. RIGHT: Subsequent evolution of the radius $R_\mathrm{sh}$, shock velocity $v_\mathrm{sh}$, and the compression ratio $\chi$ of the forward shock in direction A.}
    \label{fig:DirA}
\end{figure}

{\bf Direction A}:
This direction illustrates the characteristic features when the forward shock propagates through the bubble central region with size $\sim$ 150 pc, characterized by a relatively low density and a high temperature. The initial HD profile is shown in the left panel of Fig.\ref{fig:DirA}, where the high gas density and low $^{60}$Fe abundance at very small radii reflects the moderate-metallicity material of the stellar wind of the SN progenitor.

\begin{figure}
    \centering
    \includegraphics[width=1.0\linewidth]{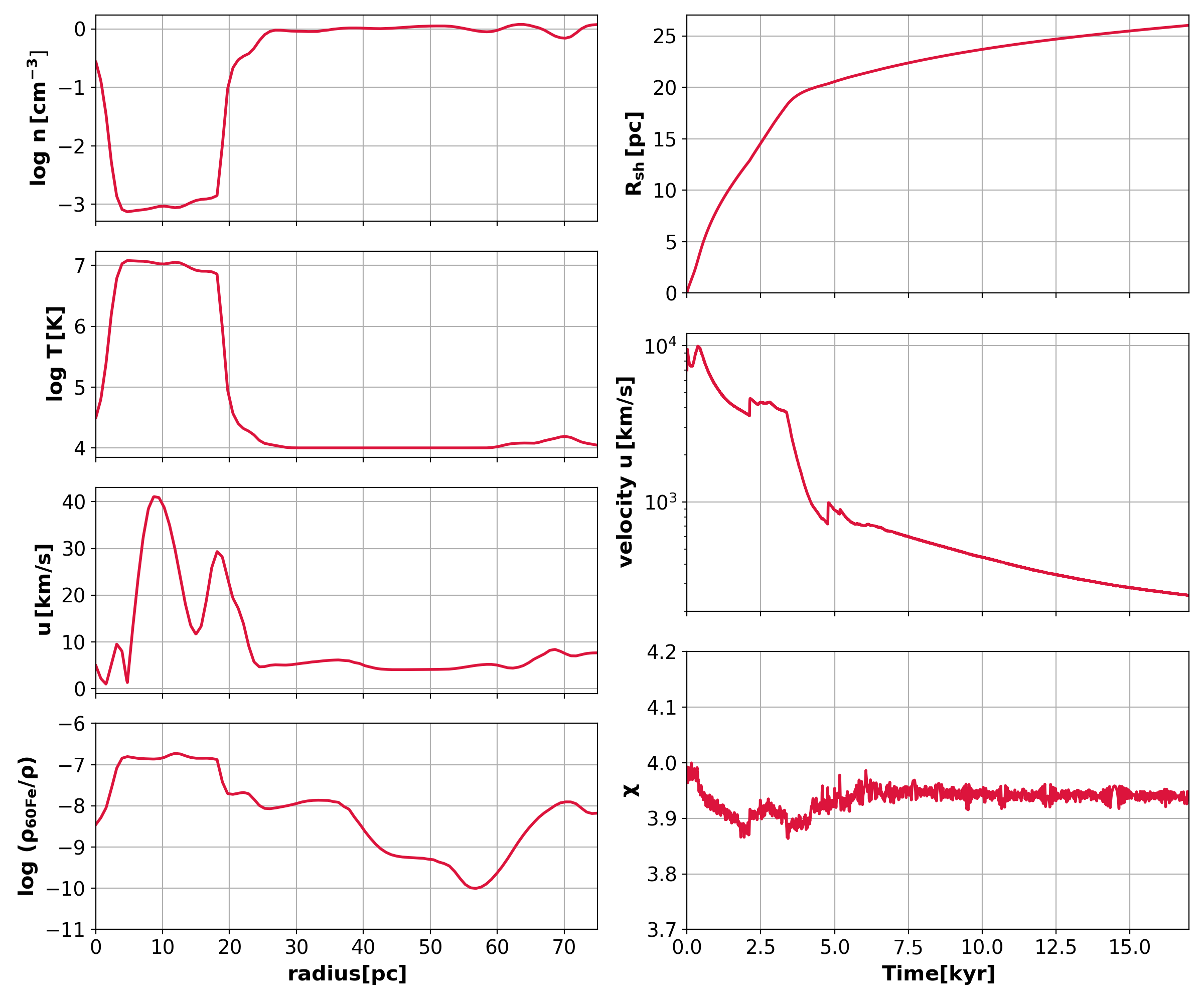}
    \caption{Initial hydrodynamic profile (LEFT) and time evolution of the forward shock (RIGHT). Same as Fig.\ref{fig:DirA} but in direction B. }
    \label{fig:DirB}
\end{figure}

The corresponding time evolution of the radius, $r_\mathrm{sh}$, speed, $v_\mathrm{sh}$, and the compression ratio $\chi$ of the forward  shock in direction A is shown in the right panel Fig.\ref{fig:DirA}.
In the initial phase, the forward shock propagates through a region of high density, picking up speed as the gas density in the environment declines. The shock speed then decreases from approximately $8,000~$km/s to around $4,000~$km/s as the shock continues to propagate, representing departure from the free-expansion phase of the SNR. 
In this phase the forward shock moves through a hot medium with moderate Mach number, causing the compression ratio to gradually decrease as well, from $\chi \approx$ 4 to 2.3.

At $t \approx 74$ kyr, the forward shock reaches the edge of the bubble, at which point the mass fraction of $^{60}$Fe drops dramatically. This interaction causes rapid deceleration, reducing the speed of the forward shock from $\approx$1,000 km/s to less than 100 km/s, with a compression ratio increasing from $\chi \approx$ 2.3 up to 3.5.

As the forward shock propagates in a complex environment, several weak reflected shocks arise that may catch up with the forward shock and accelerate it. 

{\bf Direction B}:
As the supernova progenitor resides at the edge of the LB, the forward shock is expected to hit the edge of the superbubble much earlier than it does in direction A, causing it to significantly decelerate. This early interaction with the dense shell significantly affects the resulting particle spectra. The initial HD profile is shown in the left panel of Fig \ref{fig:DirB} where, as in Figure~\ref{fig:DirA}, the wind material of the progenitor star dominates the profile at very small radial distances.

Initially, the forward shock behaves similarly as in direction A, with a significant decrease in shock speed during the free expansion phase. Throughout the entire evolution, the compression ratio of the shocks remains relatively constant at $\chi \approx 4$. After about $3,500~$years, the interaction with the dense shell of surrounding material causes further deceleration, reducing the speed of the forward shock to less than $1,000~$km/s. In this phase, the compression ratio is $\chi \simeq 3.95$. 
Then, the compression ratio remains unchanged, but shock velocity keep decreasing to $\approx$ 200 km/s at $t \approx 17$ kyr.
Because the environment is more complex than in direction A, we observe more weak reflected shocks that catch up with the forward shock and accelerate it.

\subsection{Particle spectra}

\begin{figure}
    \centering
    \includegraphics[width=1.0\linewidth]{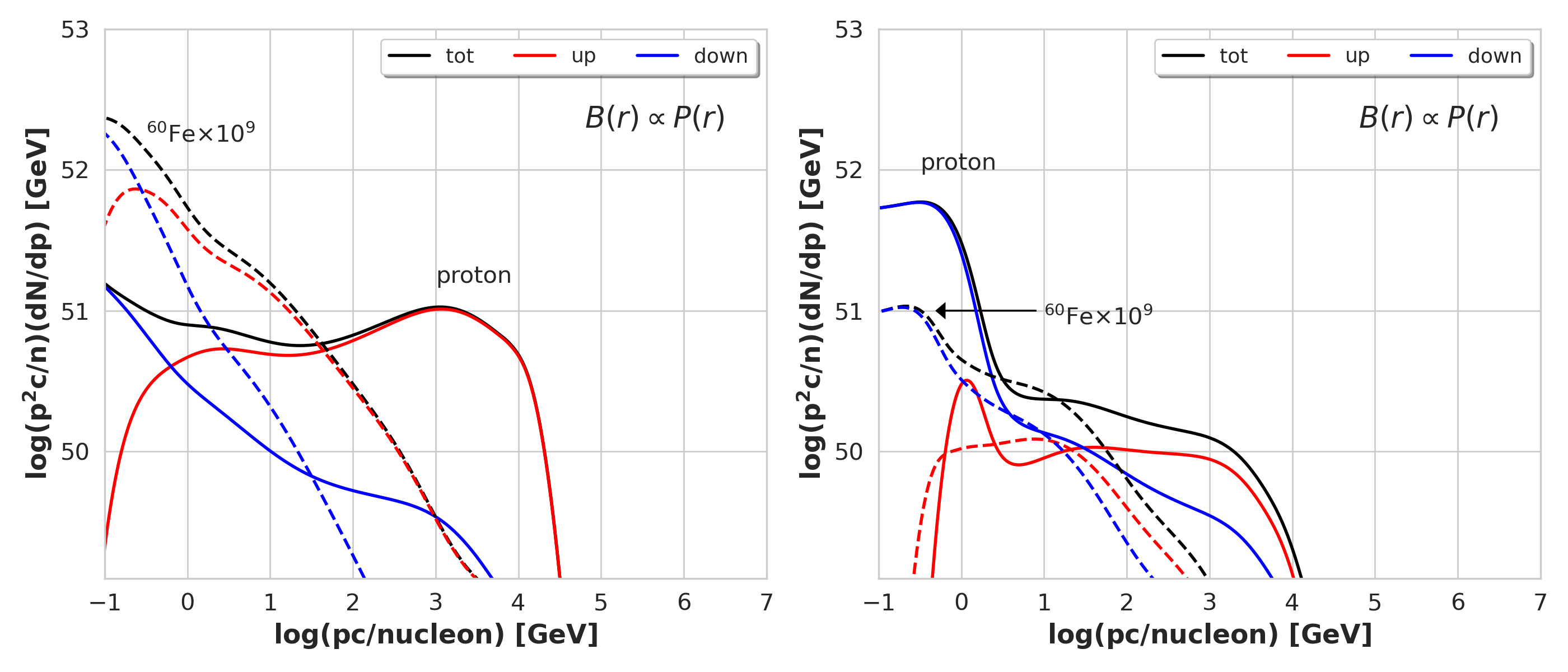}
    \caption{Proton (solid) and $^{60}$Fe (dashed) momentum spectra as functions of the momentum per nucleon at 80 kyr after the SN explosion along direction A (left) and at 17 kyr along direction B (right), both for the pressure-coupled magnetic-field profile. Black lines represent the total spectra, and blue, and red lines indicate the forward shock downstream, and the upstream spectra, respectively.}
    \label{fig:SPE_pre}
\end{figure}

To further illustrate the impact of different environmental conditions on particle acceleration for both $^{60}$Fe and protons, we present the proton and $^{60}$Fe momentum spectra in Figure \ref{fig:SPE_pre} as functions of momentum per nucleon at 80 kyr after the SN explosion along direction A and at 17 kyr along direction B, assuming the pressure-coupled magnetic-field profile. By comparing the spectra from these two directions, we can infer how varying environments influence particle acceleration and spectral shapes.

The different mass and charge of $^{60}$Fe and protons directly influence their diffusion coefficients and injection momenta.
If we assumed Bohm diffusion, the diffusion coefficients ($D \propto r_g$) would directly determine the maximum achievable momentum,  $p_\mathrm{max} \propto r_g^{-1} v_\mathrm{sh}^2$.
A key feature of our model is the disparity in spatial distributions between $^{60}$Fe and protons, which affects their injection rates across varying radii and shock speeds, leading to significant differences in the momentum spectra of these particles.

As our study employs a self-consistent treatment to calculate the diffusion coefficients, which provides a more realistic depiction of CR acceleration, the matter is more complex, on account of the interplay of the growth of Alfvén waves, wave damping, turbulence cascading, and the spatial transport of waves. Generally, particles accelerated early in the evolution are expected to reach the highest energy. At late times, most of them may reside upstream of the forward shock, because the turbulence scattering them is no longer efficiently driven, leading to poor confinement.
Most particles are accelerated at late times, when the shock has already weakened and the Mach number is low. The maximum energy reached at this late phase is very low. The spectra shown in Figure~\ref{fig:SPE_pre} reflect the total yield over the entire evolution with high maximum energy produced early on, but few particles, and more particles with lower maximum generated later, resulting in a steeper spectrum overall \citep{2020A&A...634A..59B}.

For instance, the left panel of Figure \ref{fig:SPE_pre} depicts a scenario where the forward shock propagates through a low-density, high-temperature region in direction A. 
Particle acceleration is less efficient during the very early stages of SNR evolution, when the growth rate driven by accelerated CRs is initially low.
However, as turbulence intensifies over time, the acceleration efficiency increases rapidly.
Simultaneously, the $^{60}$Fe abundance is initially low but increases significantly during the first few kyr, resulting in a distinct softer spectral.
Figure \ref{fig:SPE_pre} shows the spectra at $t = 80$ kyr after the explosion, when most of the early accelerated CRs with a hard spectrum with high maximum energy, predominantly protons, reside in the upstream region, while $^{60}$Fe exhibits a softer spectrum in the same region.
At later times, when $t = 70$ kyr, the compression ratio is $\chi \simeq 2.3$, which following the test-particle theory of diffusive shock acceleration would correspond to a theoretical spectral index of $\alpha = -(\chi+2)/(\chi-1) \simeq -3.3$, where $dN/dp \propto p^{\alpha}$.
When fitting the volume-integrated downstream spectra shown in  Figure \ref{fig:SPE_pre}, we obtain the spectral indices of $\alpha \approx -3.2$ to $-3.3$, corresponding to the momentum ranges of 10–100 GeV/nucleon for $^{60}$Fe and 100–1000 GeV/nucleon for protons. However, these are only rough estimates, as the spectra are curved and the result is sensitive to the chosen energy range.
These spectra highlight that the radial distribution of the $^{60}$Fe abundance significantly affects the $^{60}$Fe spectrum.

In direction B the forward shock's early encounter with the dense shell significantly influences particle acceleration. Generally, the high-energy spectra results from acceleration during the early stage when the shock velocity is high, which is similar to what we observe in Direction A.
The feature at lower energy appears in the later stage, when the shock velocity is lower, but the gas is dense, leading to many particles being accelerated at a lower rate and to a smaller maximum energy.

Additionally, as shown by the forward shock velocity in Figure \ref{fig:DirA} and \ref{fig:DirB} (both right panel), weak shocks catch up with the forward shock and boost it several times. These mergers increase the shock velocity and consequently enhance the acceleration efficiency of the forward shock, ultimately raising the maximum particle energy in the later stages.
Previous theoretical predictions indicate that no immediate visible change in the spectrum is expected after the merger \citep{2022ApJ...926..140S, 2022A&A...661A.128D}. Measuring the shock speed and location is difficult, when two shocks blend into each other in the simulated flow profiles, and to prevent artefacts we limit the compression ratio during the merger.

\subsection{Effects of the magnetic field profiles}

\begin{figure}
    \centering
    \includegraphics[width=\linewidth]{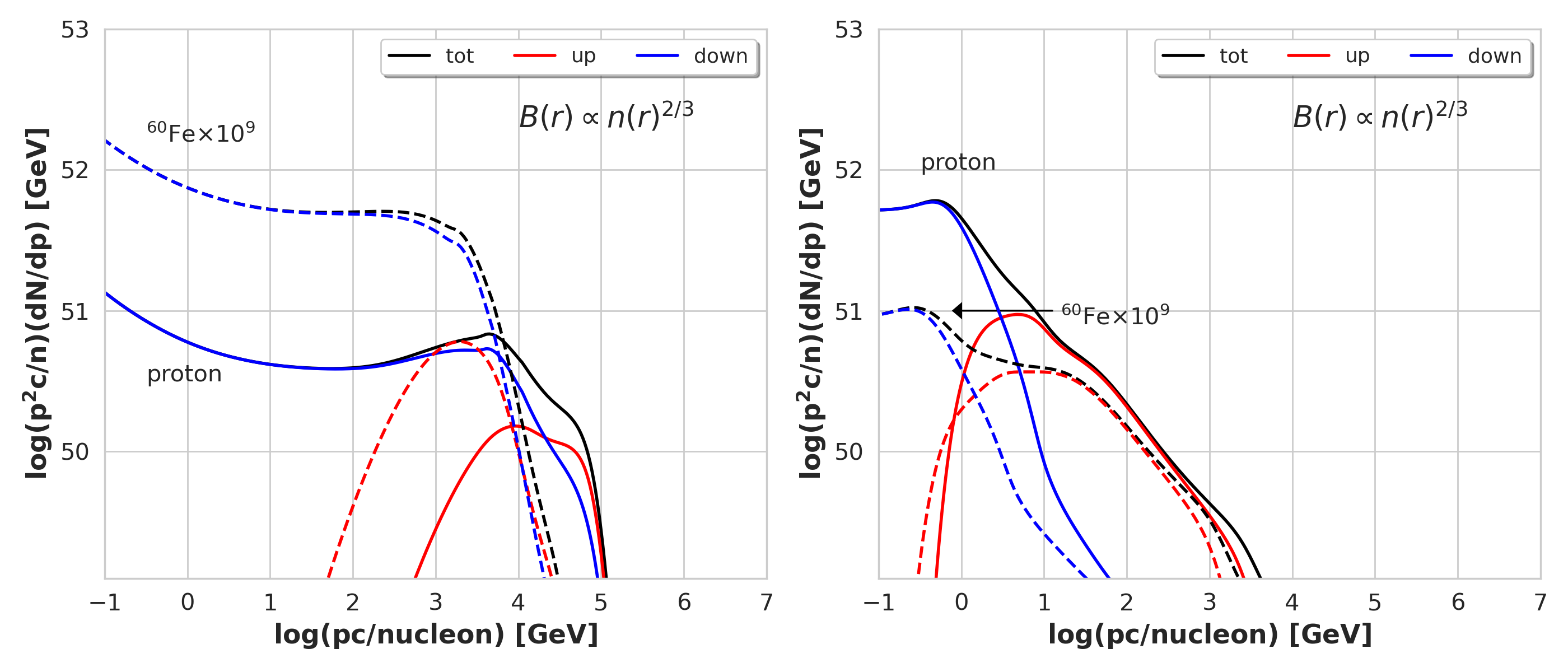}
    \caption{Same as Figure \ref{fig:SPE_pre} but under the assumption of density-coupled magnetic field profile. Proton (solid) and $^{60}$Fe (dashed) momentum spectra as function of momentum per nucleon at 80 kyr after the SN explosion along direction A (left) and at 17 kyr along direction B (right).}
    \label{fig:SPE_rho}
\end{figure}

As mentioned in Section \ref{sec:B}, we adopted two assumptions for the large-scale magnetic field in the upstream: density-coupled and pressure-coupled. In the pressure-coupled case, the magnetic field in the upstream region maintains a relatively uniform profile with $B \sim 10\ \mu$G. 

In the density-coupled case, inside the bubble region where the total density drops to $n_\mathrm{H} \lesssim 10^{-3}\ \mathrm{cm^{-3}}$, the corresponding magnetic field decreases to $B \sim 0.1\ \mu$G.
At later stages, when the forward shock progresses outward and eventually moves out of the bubble into the ISM with a typical density of $n_\mathrm{H} \sim 1\ \mathrm{cm^{-3}}$ and a magnetic field of $B \sim 10\ \mu$G, the shock velocity decreases rapidly.

When considering self-consistent diffusion, the large-scale magnetic field has influences on both the growth rate of the turbulence spectral energy density and the spatial diffusion estimation. 
These effects introduce non-linearities to the particle acceleration rates and, consequently, to the spectral shape.
As shown in \ref{fig:SPE_rho}, the relatively lower magnetic field in the bubble encouraged particle acceleration in this region, resulting in similar spectra for protons and $^{60}$Fe.

\subsection{Integrated spectra}

\begin{figure}
    \centering
    \includegraphics[width=\linewidth]{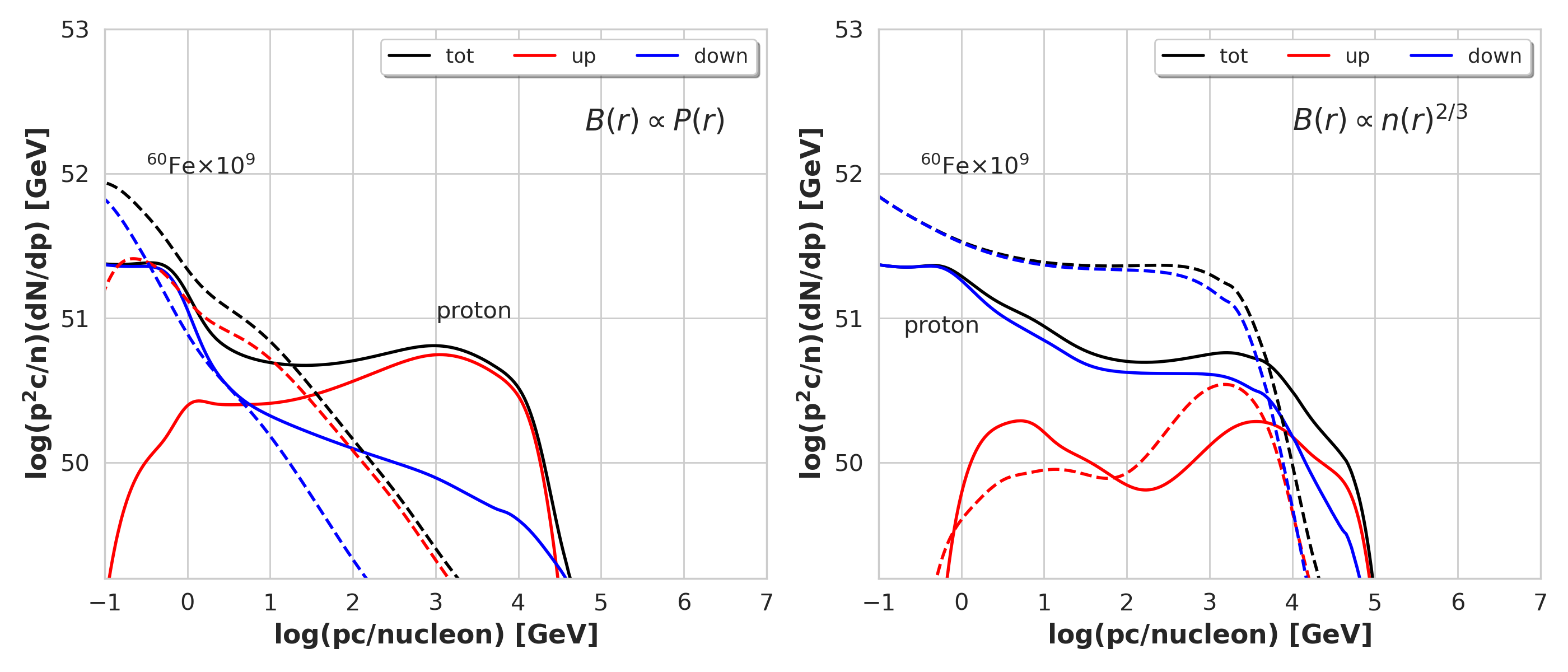}
    \caption{Integrated proton (solid) and $^{60}$Fe (dashed) momentum spectra as function of momentum per nucleon under the assumption of the pressure-coupled (left) and density-coupled (right) large-scale magnetic field profile.}
    \label{fig:int}
\end{figure}

The integrated particle momentum spectra for $^{60}$Fe and protons are averaged over 10 different directions to account for particle acceleration under varying environmental conditions. 
For each direction, the simulation is terminated when the forward shock velocity drops below 200 km/s, marking the end of efficient particle acceleration in our model.
Hence, the total simulation time varies depending on the environmental conditions along each direction. For instance, in direction A — one of the longest cases — the simulation continues for approximately 100 kyr, whereas in direction B, where the shock decelerates more rapidly, the simulation time is 20 kyr.
We then calculate the total spectra of accelerated protons and $^{60}$Fe in each direction, assuming spherical symmetry, and average them based on their spatial distribution. 
The final particle momentum spectra for $^{60}$Fe and protons are shown in Figure \ref{fig:int}.

\begin{table}[]
    \centering
    \begin{tabular}{l|c|c}
    $B(r)$  & $\propto P(r)$ & $\propto n(r)^{2/3}$ \\ \hline
    Direction A & $6.7 \times 10^{-9}$ & $1.3 \times 10^{-8}$\\
    Direction B & $1.7 \times 10^{-10}$ & $1.4 \times 10^{-10}$\\
    Integrated  & $1.5 \times 10^{-9}$ & $2 \times 10^{-9}$
    \end{tabular}
    \caption{$^{60}\text{Fe}$-to-proton number ratio in difference directions and the final integrated spectra at energy of 1~GeV/nucleon under different magnetic field assumptions.}
    \label{tab:my_label}
\end{table}

\section{Discussion}
\label{sec:dis}

We expect a higher ratio of $^{60}$Fe at low energies, as the primary acceleration sites for $^{60}$Fe are located within the bubble, leading to a softer particle spectrum.
We find
\begin{equation}
\frac{N(^{60}\mathrm{Fe})}{N(H)}\Bigg\vert_\mathrm{mod.}\simeq 2 \times 10^{-9}\  \mathrm{at}\ p_\mathrm{nuc} = 1\,\mathrm{GeV}/c ,
\label{eqmp5}
\end{equation}
corresponding to 400 MeV per nucleon in kinetic energy, fairly independent of the choice of magnetic-field scaling.
In contrast, between $10$~GeV and $10^4$~GeV per nucleon, the ratio is very much dependent on the profile of the magnetic field in the LB.

The total flux ratio of iron and protons at a momentum per nucleon of $1.3$~GeV/c is measured to be $3\times 10^{-4}$ \citep{PhysRevLett.114.171103,PhysRevLett.126.041104}. Using our prescription for particle injection (Eq.~\ref{eq:inj}) for iron and for protons in the same way and assuming ISM or solar composition with an iron abundance of $3\times 10^{-5}$ \citep{46.060407.145222}, we would find a flux ratio per CR source at $p_\mathrm{nuc} = 1\,\mathrm{GeV}$
\begin{equation}
\frac{Q(^{56}\mathrm{Fe})}{Q(\mathrm{p})}\simeq 1.5\times 10^{-5}.
\label{eqmp4}
\end{equation}
Taking the inverse and multiplying with the modelled $^{60}\text{Fe}/p$ ratio we find
\begin{equation}
\frac{Q(^{60}\mathrm{Fe})}{Q(^{56}\mathrm{Fe})}\simeq 1.3\times 10^{-4}.
\label{eqmp4a}
\end{equation}
Compared to the ISM abundance, the ion-to-proton ratio (Eq.~\ref{eqmp4})is reduced by a factor two on account of the rigidity/momentum conversion.
There is a factor of $20$ difference to the observed flux ratio of iron to protons, $3\times 10^{-4}$, that may be related to very efficient iron injection at shocks, be it by lock-up in grains or the generally larger rigidity of thermal particles compared to protons.

\citet{2016Sci...352..677B} find for local CRs a ratio $^{60}$Fe/$^{56}$Fe of $5\times 10^{-5}$, which for the $^{56}$Fe/p ratio of $1.5\times 10^{-5}$ (cf. Eq.~\ref{eqmp4}) with the iron injection scheme in our simulation would imply a flux ratio 
\begin{equation}
\frac{n(^{60}\mathrm{Fe})}{n(H)}\Bigg\vert_\mathrm{meas.}\simeq 7.5\times 10^{-10}\  \mathrm{at}\ p_\mathrm{nuc} = 1\,\mathrm{GeV}.
\label{eqmp5a}
\end{equation}
Care must be exercized in comparing these numbers with our results, because the measurement concerns a particle density, $n$, whereas we calculate the total particle count, $N$. Also, we model CRs from only one SNR, whereas the measured stable CR isotopes come from many sources and are spread over a much larger volume. For a comparison with data, we therefore need to, a), estimate the number of SNRs that contribute to the locally observed cosmic-ray flux (very few for $^{60}$Fe and very many for $^{56}$Fe) and, b), estimate the volume that is filled by these particles.

For a typical halo size $H=4$~kpc, the diffusion coefficient at $1$~GeV/n can be estimated as $D=5\times 10^{28}\ \mathrm{cm}^2\,\mathrm{s}^{-1}$ \citep{2011ApJ...729..106T}. Within time $t$ particles will fill the volume \citep{1964ocr..book.....G}
\begin{equation}
V(^{60}\mathrm{Fe})=\left(4\pi Dt\right)^{1.5}
\approx 3\ \mathrm{kpc}^3,
    \label{eqmp1}
\end{equation}
where for the last approximation we have used $1$~Myr as the time since the last supernova in the LB, a bit more than the $0.88$~Myr in the model of \citet{2023A&A...680A..39S}. It is obvious that the $^{60}$Fe yield of earlier supernovae would be diluted in a larger volume, besides the loss by radioactive decay. Most of the locally observed $^{60}$Fe flux would therefore come from the most recent supernovae in the LB.

Equation~\ref{eqmp1} most likely gives an overestimate of the volume that the recently accelerated CRs occupy. Observations of TeV haloes \citep{2020PhRvD.101j3035D,2022ApJ...936..183B} and the modeling of diffusion in the immediate environment of CR sources \citep{2020A&A...634A..59B,2021A&A...654A.139B} both indicate that the diffusion coefficient in the vicinity of CR accelerators is significantly reduced compared to that estimated for CR propagation in general.

Hydrogen nuclei and stable iron are provided by many SNRs. Given the supernova rate in the Galaxy, $Q_\mathrm{SN}\approx 0.02\ \mathrm{yr}^{-1}$, and the CR escape time, $\tau_\mathrm{esc}=H^2/2D\approx 5.5\times 10^7$~yr, one finds
\begin{equation}
    N_\mathrm{SNR, p}= Q_\mathrm{SN}\tau_\mathrm{esc}
    \approx 10^6 , \qquad N_\mathrm{SNR, Fe}= 2\times 10^5
    \label{eqmp2}
\end{equation} 
contributors to the Galactic sea of CRs. For iron nuclei, fragmentation losses reduce the lifetime by about a factor five, hence the smaller number of contributors. In either case, the particles fill a volume 
\begin{equation}
    V_\mathrm{CR}=2\pi H R_\mathrm{Gal}^2 
    \approx 5400\ \mathrm{kpc}^3, 
    \label{eqmp3}
\end{equation}
where we assumed $R_\mathrm{Gal}=15$~kpc as radius of the Galaxy. %We modelled only one SNR in the local bubble, but there are more propagated through material enriched in $^{60}$Fe. 
We can assume that propagation affects the isotopes of iron in the same way, and for the very recent production considered here decay of $^{60}$Fe is at most a 20\% effect, because the half-life at $1~$GeV/n is more than three times as long as the time since the modelled supernova. Then the flux near earth can be estimated by multiplying the yield per SNR (Eq.~\ref{eqmp4}) with the number of sources, and dividing by the volume that the particles fill. We find
\begin{equation}
\frac{n(^{60}\mathrm{Fe})}{n(^{56}\mathrm{Fe})}\Bigg\vert_\mathrm{mod.}=\frac{1}{N_\mathrm{SNR, Fe}}\,\frac{V_\mathrm{CR}}{V(^{60}\mathrm{Fe})}
\frac{Q(^{60}\mathrm{Fe})}{Q(^{56}\mathrm{Fe})}\gtrsim 1.2\times 10^{-6}, 
\label{eqmp6}
\end{equation}
which is only a few per cent of the flux ratio measured by \citet{2016Sci...352..677B}. We write this assessment as a lower limit, because we in principle need to add up the contributions of all the recent supernova in the LB, $Q/V$. That is dominated by the most recent one, on account of $V\propto t^{1.5}$ as derived in Eq.~\ref{eqmp1}, but the other supernovae in the LB may provide a moderate enhancement of the $^{60}$Fe fraction. A second source of underestimation is that the volume filled with $^{60}\mathrm{Fe}$, $V(^{60}\mathrm{Fe}) $, is very likely overestimated, as argued above.

A naive reduction of the diffusion coefficient by a factor $5$ would reduce $V(^{60}\mathrm{Fe}) $ by more than a factor of ten and would bring the estimated ratio $ \frac{n(^{60}\mathrm{Fe})}{n(^{56}\mathrm{Fe})}$ to about $30\%$ of that observed. Reality is more complex than that, and we must expect a continuous transition of the diffusion coefficient between the SNR shock and the ISM. This is a nonlinear issue in which CR transport is coupled to the acceleration efficiency at the shock. A glimpse of the complexity is offered by Fig.~\ref{fig:int}, where we see for pressure-scaling of the magnetic field in the LB that most CRs above a few GeV/c, iron and protons alike, reside upstream of the shock. For density-scaling of the magnetic field, we observed the opposite: most CRs are downstream and hence much better confined.

It is likely that additional mechanisms for $^{60}$Fe, such as the re-acceleration of the background CRs and acceleration by the reverse shock of $^{60}\mathrm{Fe}$ in the ejecta, which were not accounted for in this study, may also play a role in particle acceleration within SNRs. These mechanisms are worth further investigation in our future work.

\section{Conclusions}
\label{sec:con}
We investigated the acceleration of protons and $^{60}$Fe in an SNR in the %environment of several past SN-generated 
LB. The objective was to infer whether or not a few SN in the LB can account for the measured $^{60}$Fe flux in local CRs. Our study is based on an earlier simulation of past stellar winds and supernova activity in the LB \citep{2023A&A...680A..39S} that provides the environment in which the supernova explodes, at whose remnant we model CR acceleration and transport. 
Using RATPaC, CR transport, the large-scale gas flow (using PLUTO), as well as the large-scale magnetic field and the magnetic turbulence spectrum have been simultaneously and time-dependently computed in 1-D spherical symmetry. 
The $^{60}$Fe mass ratio is independently tracked using passive tracers.
We calculate and compare the momentum spectra of the proton and $^{60}$Fe as a function of momentum per nucleon in the LB for ten directions of expansion of the SNR. The total yield is determined as the average of the individual yields for the ten directions.

Our findings can be summarized as
\begin{itemize} 
    \item 
    The $^{60}$Fe abundance is higher inside the LB than in the dense shell and ISM, depending on direction, the sonic Mach number of the SNR shock is moderate and hence the spectrum of currently accelerated particles is soft. The modelled supernova is located at a filament of cool and dense gas, and hence the injection rate of protons is very high early and late in the evolution, when the SNR shock propagates in the ISM shell. In these regions the $^{60}$Fe abundance is small, leading to substantial differences in the spectral of ion and protons.
    \item %spectra: 
    The spectral differences between $^{60}$Fe and protons are caused by their distinct distribution in the environment of the SNR. The higher injection across the LB result in softer spectra for $^{60}$Fe, particularly in regions of low-amplitude turbulence and weaker shocks.
    \item %magnetic field: 
    The assumptions regarding large-scale magnetic fields, density-coupled or pressure-coupled, significantly influence CR acceleration. Pressure-coupled fields result in a relatively uniform magnetic profile, while density-coupled fields amplify variations, affecting the particle acceleration efficiency and the spectra.
    \item %scaling to local CR flux: 
    Our study indicates that a few supernovae in the LB can account for a fraction of the observed $^{60}$Fe flux in local CRs. The main sources of uncertainty in our estimate are, a), our modeling of only one SNR as opposed to the entire acceleration and propagation history of the past few Myrs in the LB and, b), the estimate of the volume filled with recently produced $^{60}$Fe, which primarily depends on the poorly known diffusion coefficient in the vicinity of the LB. Both theoretical arguments and the observations of TeV haloes suggest that the diffusion coefficient may indeed be significantly reduced near CR sources. Additional mechanisms such as re-acceleration of cosmic rays and acceleration of $^{60}$Fe by the reverse shock may contribute to the observed $^{60}$Fe/$^{56}$Fe ratio in CRs. 
\end{itemize}

%% Also note that the akcnowlodgment environment does not support long amounts of text. If you have a lot of people and institutions to acknowledge, do not use this command. Instead, create a new \section{Acknowledgments}.
\begin{acknowledgments}
XYS acknowledges the support by the International Postdoctoral Exchange Program of the Office of China Postdoctoral Council (OCPC) under fellow number ZD2022005.
\end{acknowledgments}


\software{Fipy \citep{2009CSE....11c...6G},  
          PLUTO \citep{2007ApJS..170..228M,2018ApJ...865..144V}, 
          RATPaC \citep{2012APh....35..300T, 2013A&A...552A.102T, 2018A&A...618A.155S, 2020A&A...634A..59B}
          }


\bibliography{fe60}{}
\bibliographystyle{aasjournal}



\end{document}